\documentclass{webofc}
\usepackage[varg]{txfonts}   
\usepackage{graphicx}
\usepackage{bm}
\usepackage{epstopdf}

\woctitle{10th International Workshop on Multiple Partonic Interactions at the LHC}

\def\bea{\begin{eqnarray}}
\def\eea{\end{eqnarray}}

\def\pp{\mbox{$p$-$p$}}
\def\pa{\mbox{$p$-$A$}}

\def\pbpb{\mbox{Pb-Pb}}
\def\aa{\mbox{$A$-$A$}}
\def\nn{\mbox{$N$-$N$}}

\def\ee{\mbox{$e^+$-$e^-$}}
\def\ppbar{\mbox{$p$-$\bar p$}}

\def\pt{$p_t$}
\def\yt{$y_t$}
\def\nch{$n_{ch}$}
\def\v2{$v_2$}
\def\mmpt{$\bar p_t$}
\def\ppb{\mbox{$p$-Pb}}
\def\pn{\mbox{$p$-$N$}}

\begin{document}
\title{PYTHIA and the preoccupied proton}

\author{Thomas A. Trainor\inst{1}\thanks{\email{ttrainor99@gmail.com}}
}

\institute{University of Washington, Seattle, Washington, USA
}

\abstract{%
The PYTHIA Monte Carlo (PMC) has been applied broadly to simulations of high-energy $p$-$p$ and $p$-$\bar p$ collisions. 
The PMC is based on several assumptions, such as that most hadrons result from jet production (multiple parton interactions or MPIs), that $p$-$p$ centrality is relevant and that color reconnection (CR) strongly influences fragmentation to jets.  An alternative description is provided by the two-component (soft + hard) model (TCM) of hadron production. TCM analysis of $p$-Pb ensemble-mean-$p_t$ data reveals centrality trends quite different from those estimated via a geometric Glauber model based on the eikonal approximation. 
Glauber estimates of binary-collision number are  three times TCM estimates. Detailed study of $p$-Pb data conflicts with a basic Glauber assumption -- that a projectile proton may interact simultaneously with multiple target nucleons. Instead, in both $p$-$p$ and $p$-A collisions, a $p$-N collision once initiated is exclusive of other possible interactions (during that collision), and within the collision any pair of participant partons may interact -- a $p$-N collision is thus ``all or nothing.'' In this presentation the PMC is challenged by an assortment of contradictory data, and evidence for $p$-N exclusivity is reviewed
 to make a case for the ``preoccupied proton'' of the title. 
}
\maketitle
\section{Introduction}
\label{intro}

This presentation confronts the PYTHIA Monte Carlo (PMC) with an alternative two-component (soft + hard) model (TCM) of hadron production near midrapidity. Modeling of \pp\ centrality is a major issue. Centrality modeling for \pa\ collisions provides important evidence against the relevance of centrality in elementary \nn\ collisions. Results suggest that a new principle of {\em exclusivity} governs  binary \nn\ collisions within composite A-B collisions, leading to the ``preoccupied proton'' of the title.

Within the PMC multiple parton interactions (MPIs) are assumed to be the dominant or exclusive mechanism for hadron production.  The PMC thus contains no feature comparable to the TCM soft component.  Non-diffractive inelastic scattering is modeled within the PMC by extending the pQCD parton scattering cross section down to \pt\ = 0 with collision-energy-dependent soft cutoff parameter $p_{\perp0}$ as a main tuning parameter. \pp\ centrality described by the Glauber model with eikonal approximation is a major feature of the model. Color reconnection (CR) is assumed so as to minimize the total string length (i.e.\ fragment number) resulting from multiple hard parton scatters (MPIs). The PMC is tuned to accommodate a specific subset of currently available data volumes and analysis methods~\cite{sjostrand}. 

In contrast, the TCM is interpreted such that for hadron production near midrapidity the soft component represents the majority of hadrons, those arising from longitudinal dissociation or fragmentation of projectile nucleons, while the minority hard component represents hadrons from large-angle-scattered partons (gluons) fragmenting to minimum-bias (MB) dijets~\cite{ppprd,ppquad,tommpt}. The observation in \pp\ collisions of a quadratic relation between soft and hard components~\cite{ppprd} is a major result with its own implications for \pp\ centrality as modeled in the PMC. The TCM provides an accurate description of a broad array of data including \mmpt\ data for \ppb\ collisions as described in this presentation~\cite{tompythia}. 

\section{Two-component model of hadron production}

The TCM was first applied to hadron production in \pp\ (\ppbar) collisions near midrapidity~\cite{ppprd,pancheri}. Mean charge density $\bar \rho_0 = n_{ch} / \Delta \eta = \ \bar \rho_s + \bar \rho_h$ averaged over some $\eta$ acceptance $\Delta \eta$ is decomposed into soft and hard components.  Soft-component charge density $\bar \rho_s = n_s / \Delta \eta$ is a product of projectile nucleon dissociation along $z$ and corresponds to participant low-$x$ gluons, with $\bar \rho_s \propto \log(\sqrt{s} / \text{10 GeV})$. Hard-component density $\bar \rho_h$ consists of fragments from MB dijets. Those interpretations are supported by comparing TCM results for a variety of collision systems to measured jet properties~\cite{hardspec,fragevo,alicetomspec}. An essential feature of the \pp\ TCM is the observed quadratic relation $\bar \rho_h \approx \alpha \bar \rho_s^2$ with $\alpha \approx O(0.01)$. Generalized to composite A-B systems (A-B geometry parameters and \nn\ densities are factorized) the spectrum TCM is

\bea \label{spectcm}
\bar \rho_0(y_t) &=& (N_{part} / 2)\, \bar \rho_{sNN} \hat S_0(y_t) + N_{bin}\, \bar \rho_{hNN} \hat H_0(y_t)
\\ \nonumber
\frac{\bar \rho_0(y_t)}{\bar \rho_s} &=& \hat S_0(y_t) + x(n_s) \nu(n_s) \hat H_0(y_t),
\eea
where $y_t \equiv \ln[(m_t + p_t) / m_\pi]$, $x(n_s) = \bar \rho_{hNN} / \bar \rho_{sNN} \approx \alpha \bar \rho_{sNN}$ and $\nu \equiv 2 N_{bin} / N_{part}$. The hatted soft and hard model functions on \yt\ are unit normal.

\section{Minimum-bias dijets, the TCM hard component and MPIs} \label{sec-1}

The PMC is based in part on the assumption that most hadrons from a \pp\ collision emerge from multiparton interactions (per event) or MPIs (i.e.\ jets) described by pQCD. The PMC is then in effect a one-component model or OCM (hard only). However, experimental evidence conflicts with that assumption~\cite{ppprd,ppquad,hardspec,fragevo,alicetomspec}.

\begin{figure}[h]
	\includegraphics[width=1.23in]{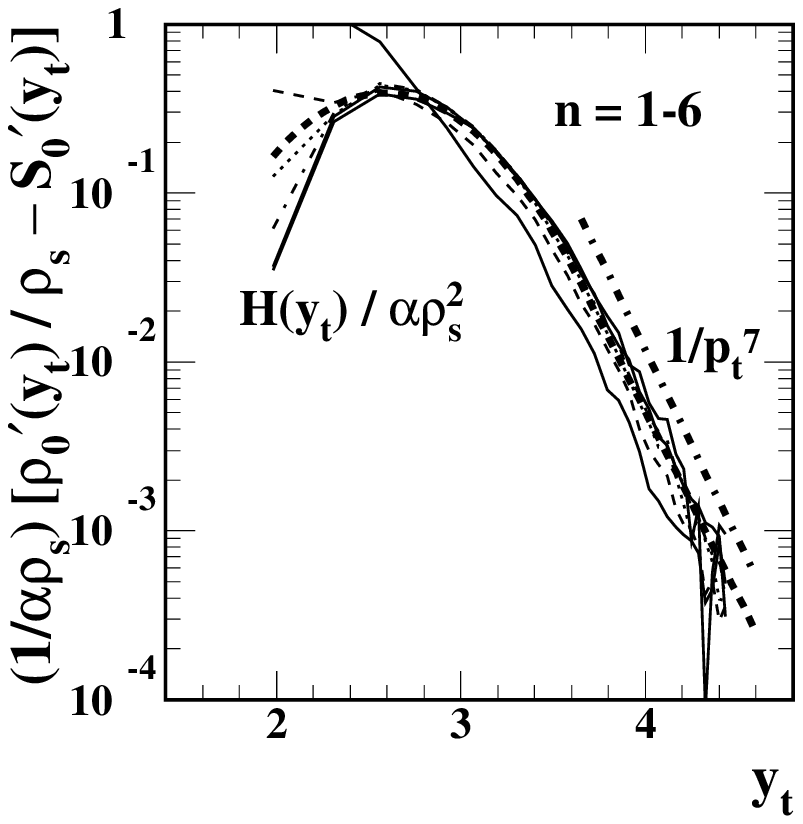}
	\includegraphics[width=1.26in]{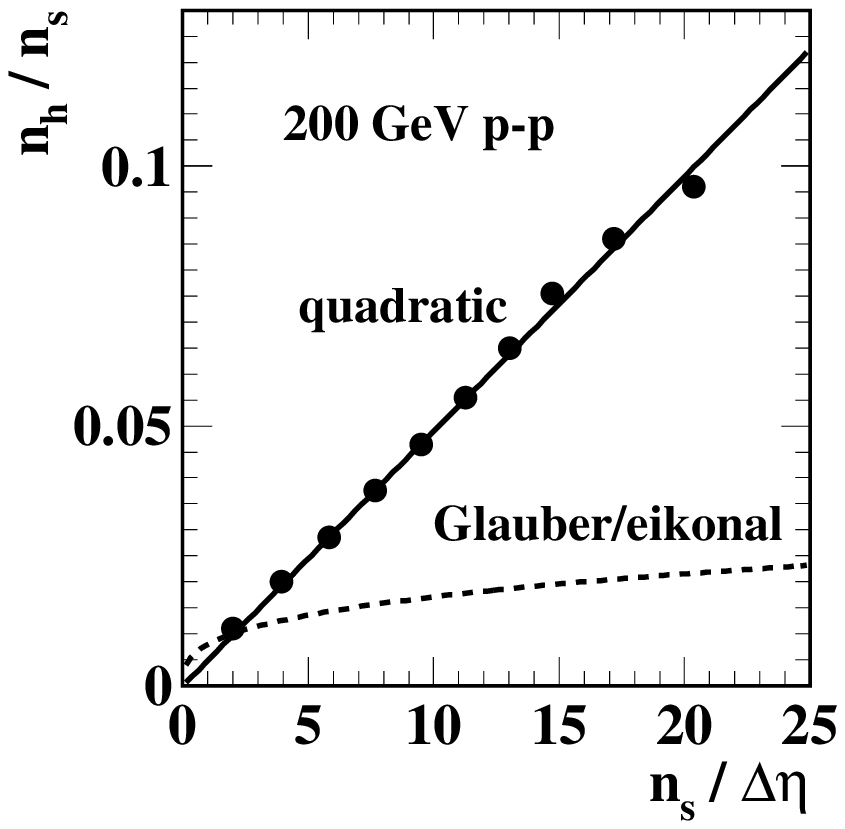}
	\includegraphics[width=1.25in]{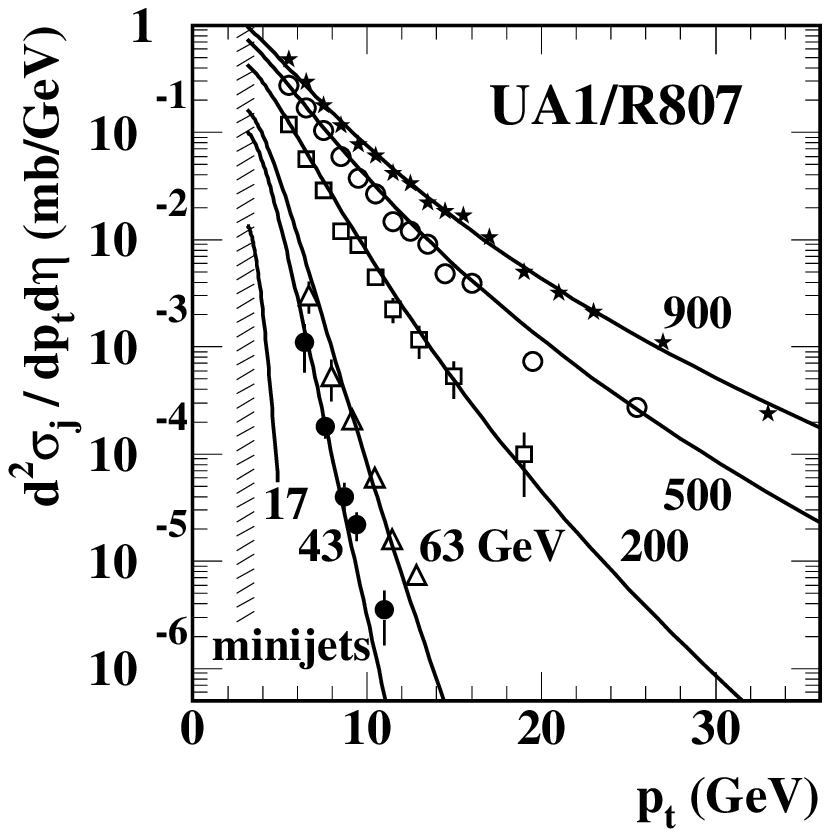}
	\includegraphics[width=1.25in]{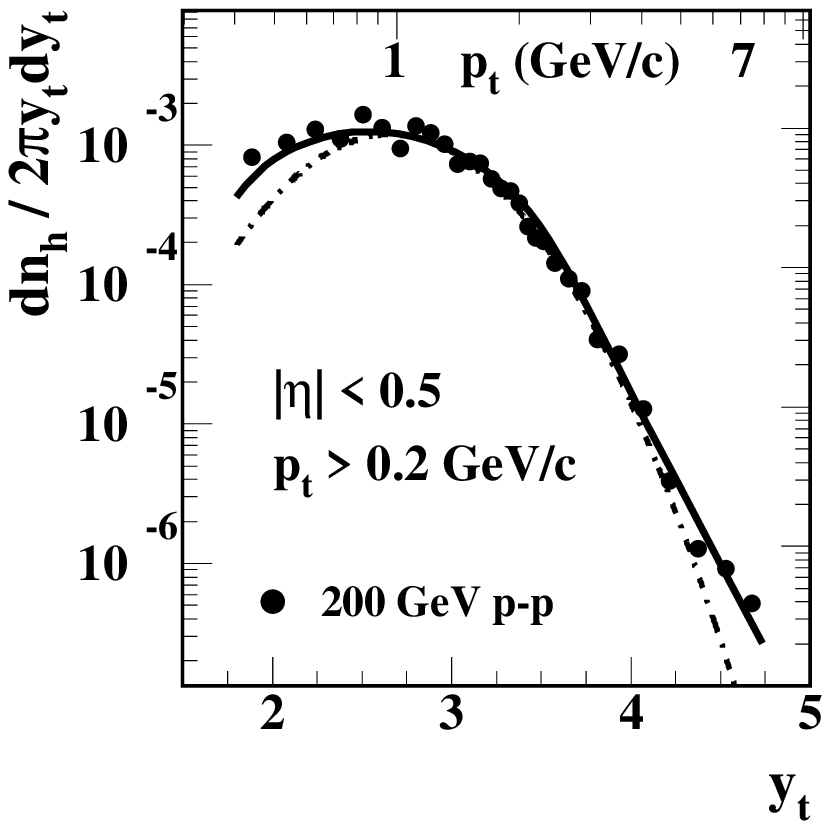}
	\caption{\label{fig1}
First: Spectrum hard components for six multiplicity classes of 200 GeV \pp\ collisions.
Second: Relation between soft and hard charge yields.
Third: Jet spectra for five \pp\ collision energies (points) and associated  curves from Eq.~(\ref{jetspec}).
Fourth: Hard component for NSD \pp\ collisions (points) and pQCD prediction from jet measurements (solid curve).
	}   
\end{figure}

Figure~\ref{fig1} (first) shows hadron spectrum hard components for six multiplicity \nch\ classes of 200 GeV \pp\ collisions (thin curves of several line styles). The bold dashed curve is hard-component model $\hat H_0(y_t)$, a Gaussian with exponential tail. The systematic deviations from the model have been interpreted to result from bias of the underlying jet spectrum by the imposed \nch\ condition~\cite{alicetomspec}. The second panel shows the ratio $n_h / n_s$ of soft and hard yields integrated within acceptance $\Delta \eta = 1$ vs $\bar \rho_s$. The data follow a linear trend accurately corresponding to $\bar \rho_h \approx \alpha \bar \rho_s^2$ over a ten-fold increase in $\bar \rho_s$, corresponding therefore to a {\em 100-fold increase in dijet production}~\cite{ppquad}. The first panel indicates no significant alteration of jet formation over that interval, in conflict with color reconnection as described below. 

Figure~\ref{fig1} (third) shows jet \pt\ spectra (points) from \pp\ (\ppbar) collisions at five collision energies. The solid curves are determined by the model function~\cite{jetspec2}
\bea \label{jetspec}
p_t\frac{d^2\sigma_j}{dp_t d\eta} &=& \frac{d^2\sigma_j}{dy_{max} d\eta} = 0.052 \Delta y_b^2 \frac{1}{2\pi \sigma_u^2}e^{-u^2 / 2 \sigma_u^2}, ~~~~~\sigma_u \approx 1/7
\eea
where $y_{max} = \ln(p_t / m_\pi)$, $\Delta y_{max} = \ln(\sqrt{s} / 2E_{cut})$, $u = \ln(p_t / E_{cut}) / \Delta y_{max} \in [0,1]$ and $\Delta y_b = \ln(\sqrt{s} / \text{10 GeV}) \propto \bar \rho_s$, with $E_{cut} \approx 3$ GeV being the effective lower bound of a jet spectrum. Factor $\Delta y_b^2$ corresponds exactly to the quadratic relation between soft and hard components in Eq.~(\ref{spectcm}). The exponential is a Gaussian on normalized rapidity $u$. Jet spectra for all collision energies are described by that simple function.

Figure~\ref{fig1} (fourth) shows the spectrum hard component for 200 GeV NSD \pp\ collisions (points) as in the first panel. The solid curve is a convolution of the 200 GeV jet spectrum in the third panel and CDF-measured \ppbar\ fragmentation functions~\cite{fragevo}. A lower bound of 3 GeV on the jet spectrum is required by the hadron spectrum data. Given those results the integral of Eq.~(\ref{jetspec}) over \pt\ and $\eta$ estimates a jet total cross section $\sigma_{j0} \approx 4$ mb which can be compared with the NSD cross section 36 mb for that collision energy to infer that the frequency per event of MB dijets in $4\pi$ is $O(0.1)$ at 200 GeV.

\section{PYTHIA and the underlying event -- UE} \label{sec-1}

In the context of the PMC a triggered dijet is assumed to be confined to two regions on azimuth centered at 0 and $\pi$ relative to the trigger. Complementary regions centered at $\pm \pi/2$ are described as the transverse region or TR. The integrated charge within the TR is denoted by $N_\perp$, and the trend of $N_\perp$ with trigger \pt\ condition is said to reveal properties of the {\em underlying event} or UE. In particular, the rise of $N_\perp$ with $p_t$(jet) to a saturation value (pedestal) should reflect increase of the MPI rate with increasing \pp\ centrality in response to the trigger condition~\cite{sjostrand,frank}.

\begin{figure}[h]
	\includegraphics[width=1.25in,height=1.32in]{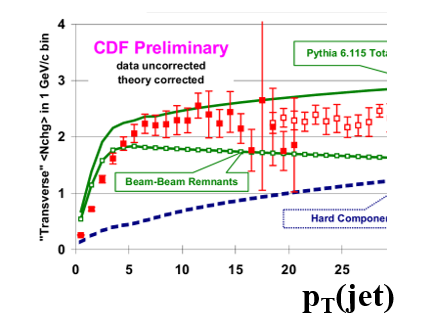}
	\includegraphics[width=1.25in]{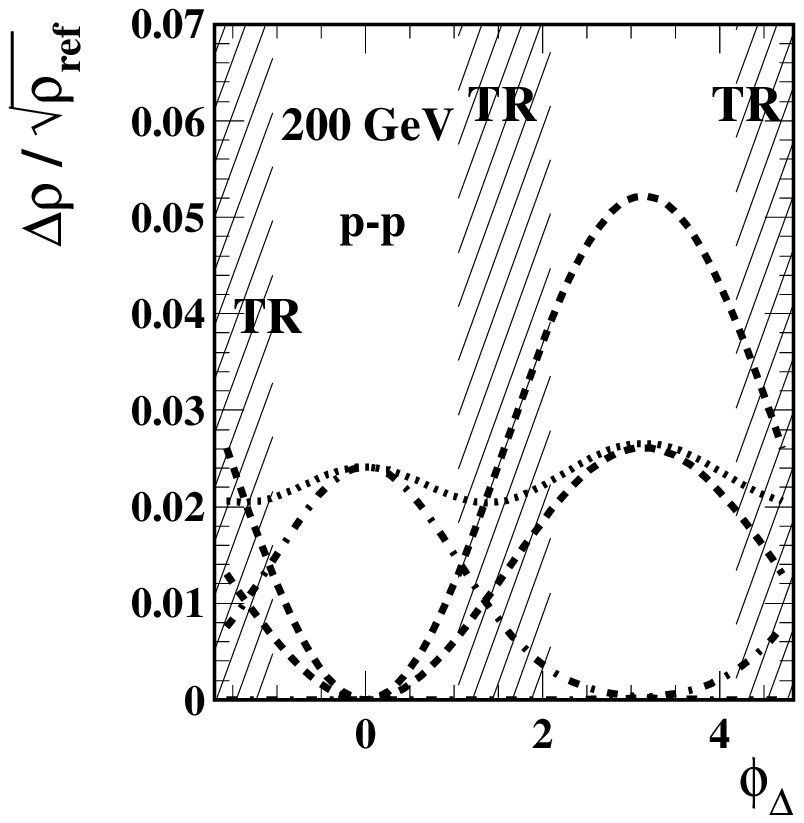}
	\includegraphics[width=1.25in]{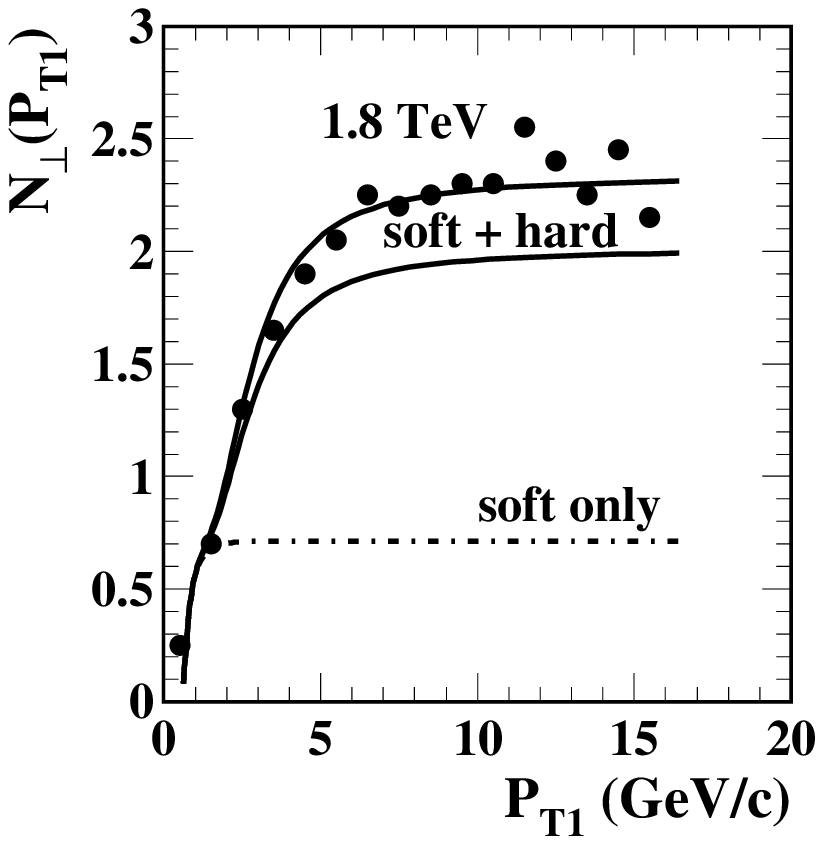}
	\includegraphics[width=1.25in,height=1.25in]{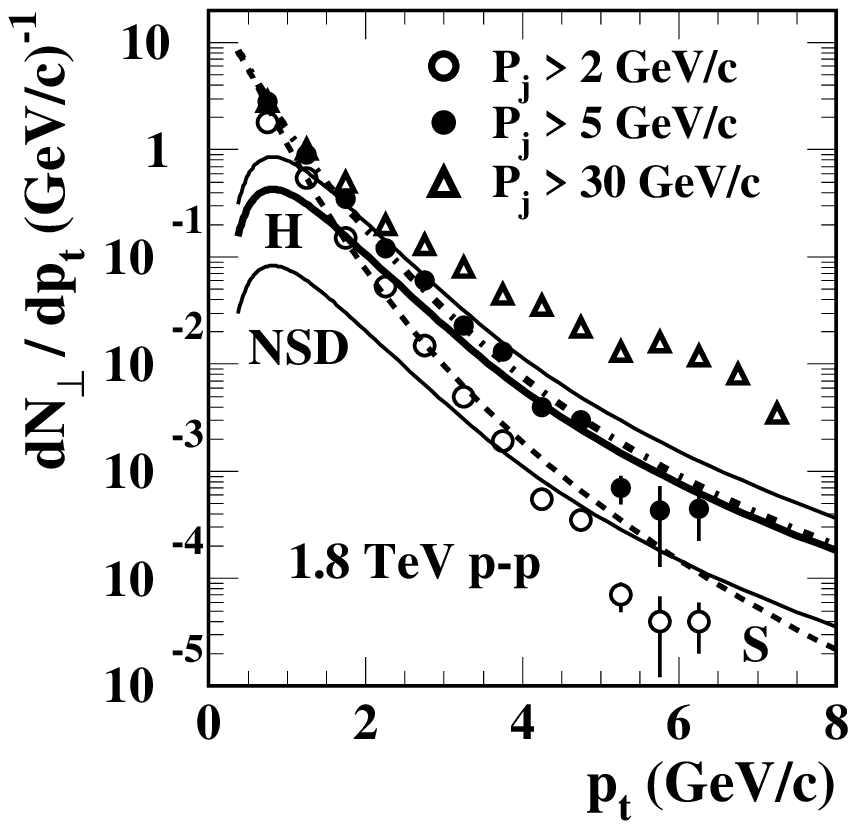}
	\caption{\label{fig2}
First: CDF data associated with the underlying event (points) compared to PYTHIA predictions (curves).
Second: Model functions derived from 2D angular correlations.
Third: Data from first panel (points) compared to running integrals of single-particle spectra (solid curves).
Fourth: $N_\perp$ spectra (points) compared to TCM predictions (curves).
	}   
\end{figure}

Figure~\ref{fig2} (first) shows CDF UE data for $N_\perp$ vs \pt(jet) (points) in a conventional plot format~\cite{under5}. The curves are from PYTHIA. The second panel shows the fragment azimuth distribution for MB dijets from 200 GeV \pp\ collisions (dotted curve) derived from model fits to 2D angular correlations~\cite{ppquad}. The relevant conclusion from this plot is that any dijet, triggered or not, must make a substantial contribution to the TR.

Figure~\ref{fig2} (third) shows $N_\perp$ data from the first panel (points) compared to running integrals of the TCM for 1.8 TeV \ppbar\ \pt\ spectra (dash-dotted and lower solid)~\cite{pptheory}. Those curves are TCM predictions based on the spectrum TCM reported in Ref.~\cite{alicetomspec}, not fits to data. The panel demonstrates that the UE pedestal effect does not result from increasing \pp\ centrality, instead reflects running integration of the single-particle \pt\ spectrum in response to the jet \pt\ trigger condition. It is notable that because of the 0.5 GeV/c \pt\ acceptance lower limit for the CDF data only 25\% of the spectrum soft component is accepted. With full \pt\ acceptance the soft component would dominate the $N_\perp$ data.

Figure~\ref{fig2} (fourth) shows  $N_\perp$ \pt\ spectra for three trigger conditions (points). The curves represent a TCM prediction for 1.8 TeV derived from the spectrum TCM in Ref.~\cite{alicetomspec}, the dashed curve being the fixed soft component. The condition $P_j > 2$ GeV/c completely suppresses jet production; data are consistent with the soft component alone. Condition $P_j > 5$ GeV/c corresponds to 50\% of events with a dijet (bold solid), and  $P_j > 30$ GeV/c corresponds to 100\% (upper thin solid). For any trigger condition the data at low \pt\ are consistent with the fixed soft component alone, arguably representing the ``underlying event.'' There is no variation corresponding to changing \pp\ centrality. Data at higher \pt\ are  biased by the trigger condition as one might expect if the triggered jet makes a strong contribution to the TR. The same trigger bias corresponds to the upper solid curve in the third panel (with 20\% increase of the TCM hard component to accommodate the data).

\section{PYTHIA and color reconnection -- CR} \label{sec-1}

The default PMC is unable to reproduce \pp\ \mmpt\ vs \nch\ data; to accommodate such data a {\em color reconnection} or CR mechanism was introduced~\cite{sjostrand}. Color connections (e.g.\ strings) said to result from multiple parton interactions (MPIs) are rearranged or reconnected to minimize total string length. A consequence of such rearrangement should be a resulting strong dependence of jet formation on jet number or density.

\begin{figure}[h]
	\includegraphics[width=1.25in,height=1.26in]{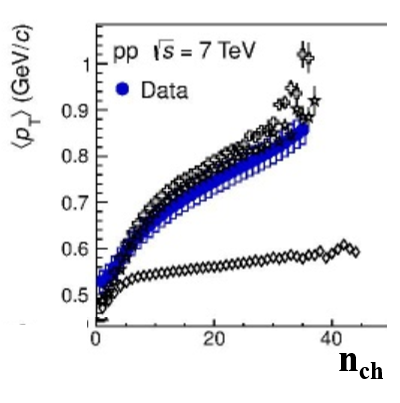}
	\includegraphics[width=1.25in]{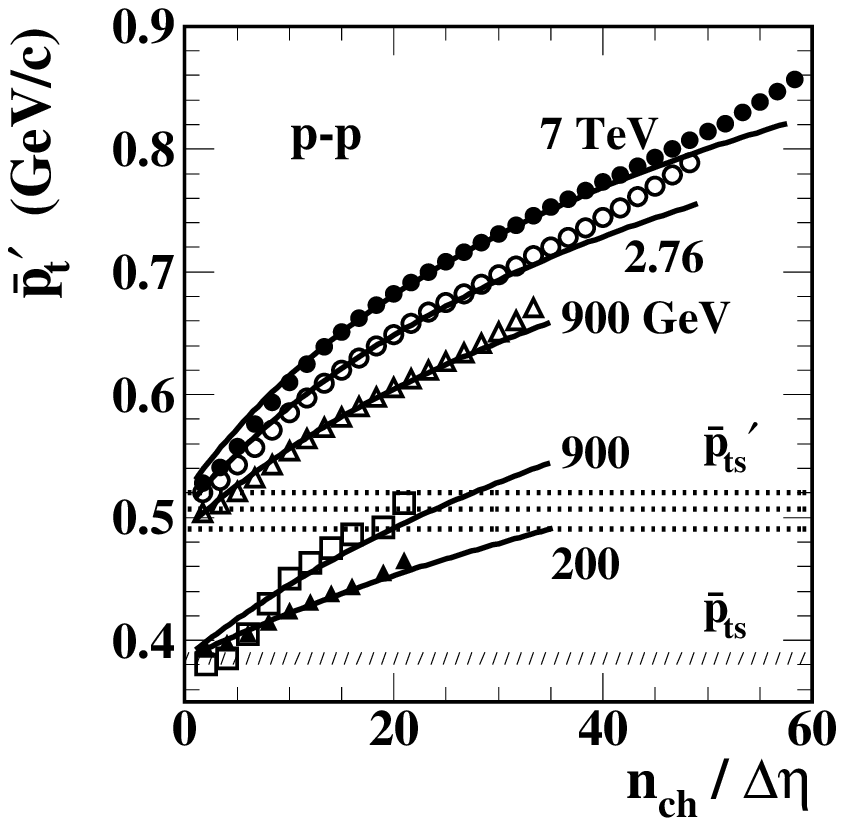}
	\includegraphics[width=1.2in]{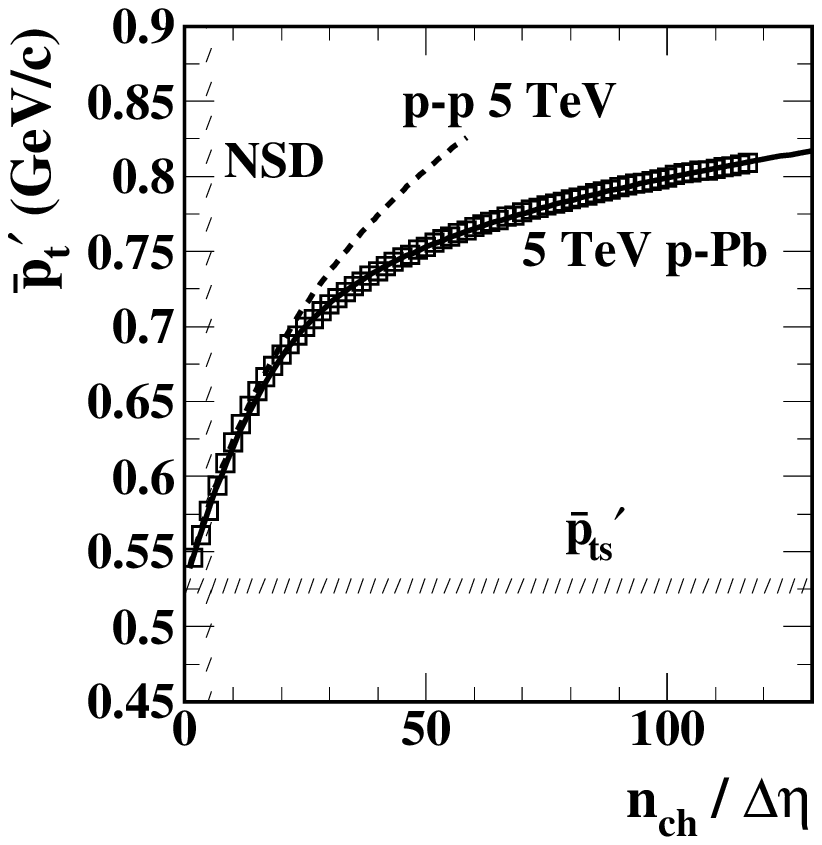}
	\includegraphics[width=1.2in]{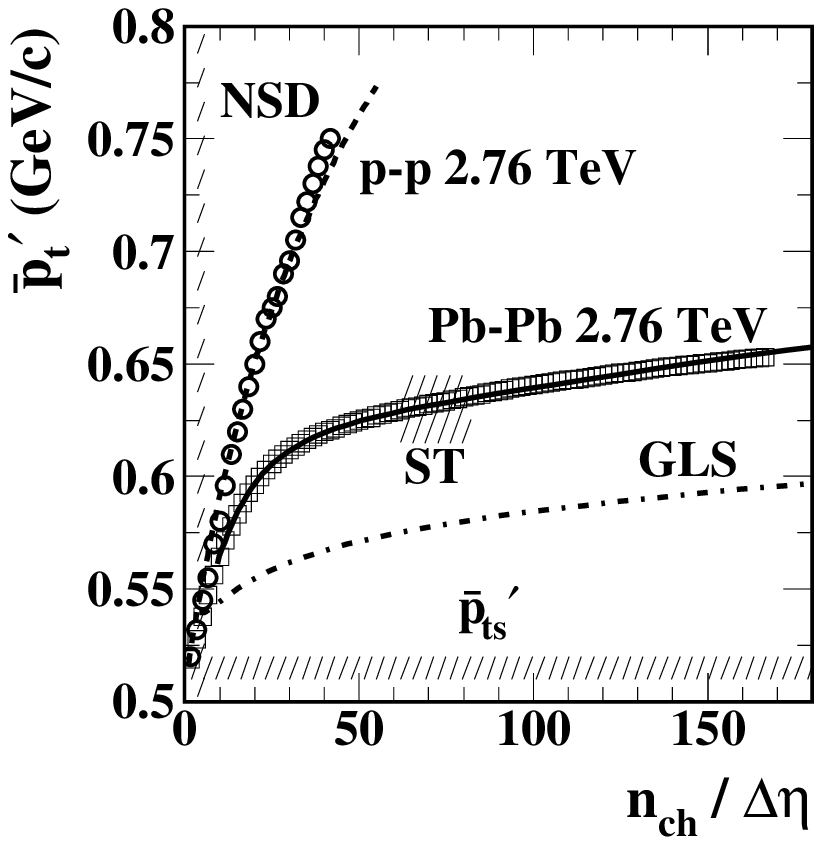}
	\caption{\label{fig3}
		First: ALICE \mmpt\ data (solid points) compared to PYTHIA predictions (open points).
		Second:  \mmpt\ data (points) compared to TCM trends (curves).
		Third: ALICE \mmpt\ data for 5 TeV \ppb\ collisions (points) compared to a TCM trend (solid curve).
		Fourth: ALICE data for 2.76 TeV \pbpb\ collisions (open boxes) compared to a TCM trend (solid curve).
	}   
\end{figure}

Figure~\ref{fig3} (first) shows  \mmpt\ vs \nch\ (\nch\ integrated within  $\Delta \eta = 0.6$) data (solid points) for 7 TeV \pp\ collisions from Ref.~\cite{alicempt}. The lower open points from the default PMC can be compared with the curve marked ``Glauber/eikonal'' in Fig.~\ref{fig1} (second). The upper open points are from the PMC with CR mechanism adjusted to accommodate the data. The second panel includes those 7 TeV \pp\ data and \mmpt\ data from other collision energies (points). The curves are derived from a comprehensive \pt\ spectrum TCM that accurately describes spectrum data from SPS to top LHC energies~\cite{alicetomspec,tommpt}. Data deviations from the TCM curves result from small shifts of hard components on \yt\ with varying \nch~\cite{alicetomspec}. The accurate TCM representation of \pp\ \mmpt\ data without a CR mechanism arises from the quadratic relation $\bar \rho_h \propto \bar \rho_s^2$  also reflected in the jet spectrum description of Eq.~\ref{jetspec}.

Figure~\ref{fig3} (third) shows \mmpt\ data for 5 TeV \ppb\ collisions (points) compared to the TCM reported in Ref.~\cite{tommpt} (solid). The dashed curve represents the \pp\ TCM as in the second panel but with interpolation to 5 TeV. The data are described within their uncertainties. It is notable that \pp\ and \ppb\ data coincide precisely up to $\bar \rho_0 \approx 20$. The fourth panel shows a comparable TCM description of 2.76 TeV \pbpb\ \mmpt\ data.

\section{$\bf p$-Pb centrality analysis: TCM vs Glauber model} \label{sec-1}

\ppb\ centrality determination can be inferred from the TCM for \mmpt\ data as described in Refs.~\cite{tommpt,tomglauber}. The TCM for extensive mean {\em total} \pt\ integrated over all \yt\ and some $\Delta \eta$ acceptance is derived from Eqs.~\ref{spectcm}
\bea
\bar P_t &=& (N_{part} / 2)\, n_{sNN} \bar p_{tsNN} + N_{bin}\, n_{hNN}  \bar p_{thNN}(n_s)
\\ \nonumber
\frac{\bar P_t}{n_s} &=& \bar p_{ts} + x(n_s) \nu(n_s)  \bar p_{th0},
\eea
where universal $\bar p_{ts} \approx 0.40$ GeV/c corresponds to universal slope parameter $T \approx 145$ MeV within $\hat S_0(y_t)$.
The second line assumes $\bar p_{tsNN}(n_s) \rightarrow  \bar p_{th0}$ fixed, i.e.\ no jet modification in \ppb\ collisions. The two $\bar p_{tx}$ values are determined by model functions $\hat S_0(y_t)$ and $\hat H_0(y_t)$. Hard/soft ratio  $x(n_s) = n_{hNN} / n_{sNN} \approx \alpha \bar \rho_{sNN}$ is assumed as in \pp\ collisions (Fig.~\ref{fig1}, second). If  $x(n_s)$ is defined other parts of the TCM are also determined as a consequence.

\begin{figure}[h]
	\includegraphics[width=1.25in]{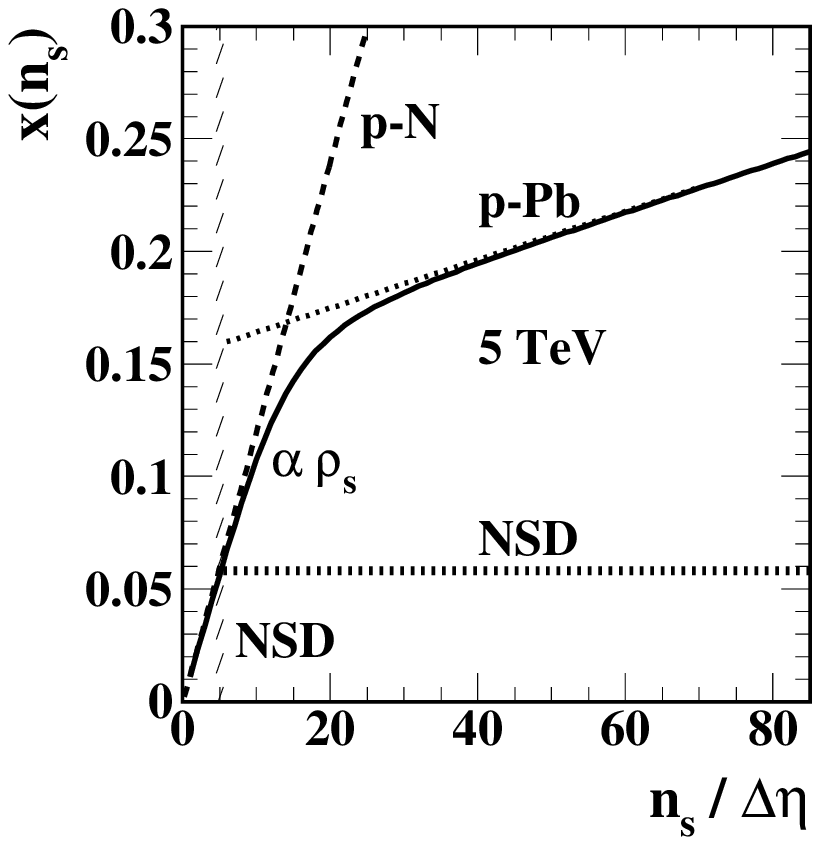}
	\includegraphics[width=1.25in]{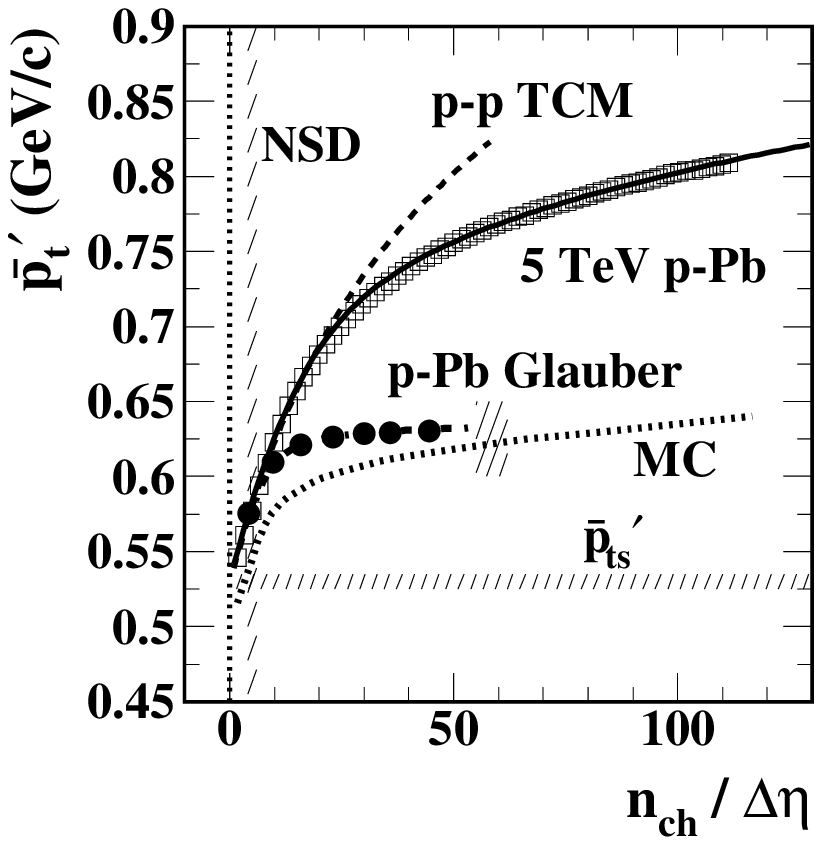}
	\includegraphics[width=1.25in]{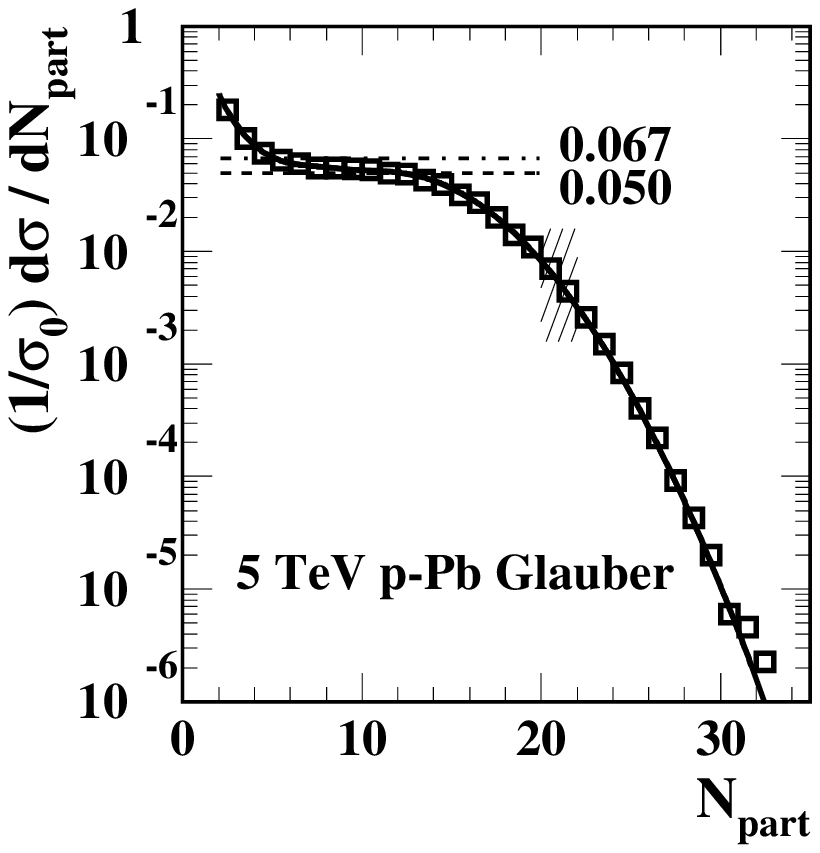}
\includegraphics[width=1.25in]{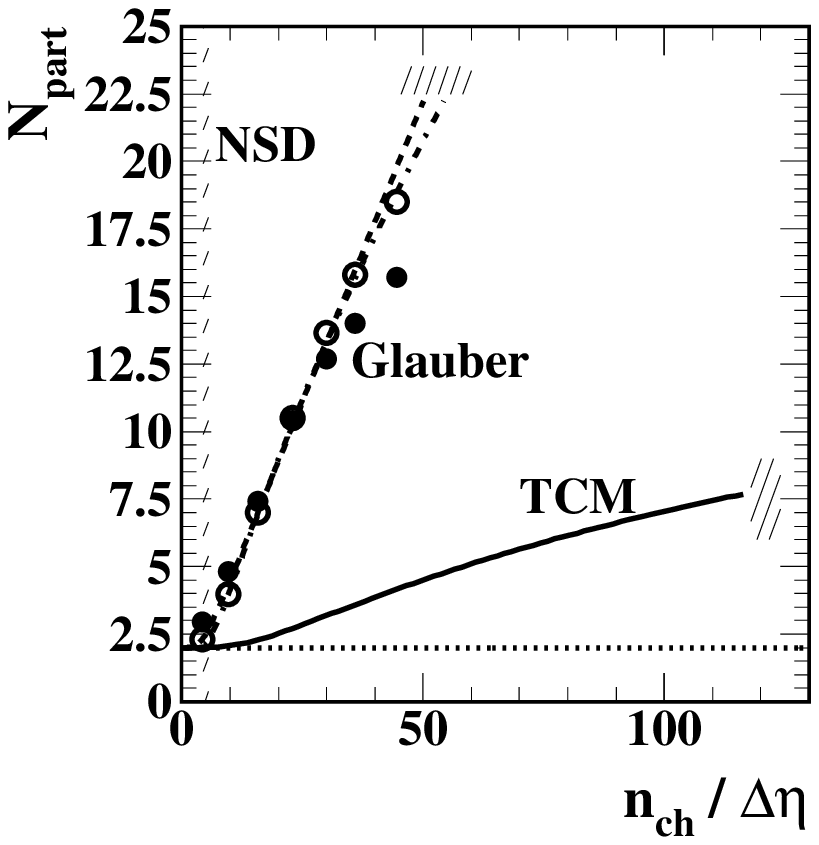}
	\caption{\label{fig4}
First: Model for TCM $x(n_s)$ parameter (solid curve).
Second: 5 TeV \ppb\ \mmpt\ data (open points) compared to TCM (solid curve) and predictions from a Glauber analysis (solid points).
Third: Glauber-model cross-section distribution on $N_{part}$.
Fourth: Comparison of Glauber (points) and TCM (solid curve) predictions  for $N_{part}$ vs charge density $\bar \rho_0$.
	}   
\end{figure}

Figure~\ref{fig4} (first) shows $x(n_s)$ vs $\bar \rho_s$ for \ppb\ collisions (solid curve) based on agreement between \pp\ and \ppb\ \mmpt\ data below $\bar \rho_0 \approx 20$ (dashed line) and the simplest possible modification above that point, a linear trend with reduced slope (dotted line). The second panel shows the resulting TCM (solid curve) compared to \mmpt\ data (open squares). The data are described within their uncertainties. Transition point $\bar \rho_{s0} \approx 15$ and slope reduction factor $m_0 \approx 0.1$ are the only new parameters required to extend the \pp\ TCM to \ppb\ data~\cite{tommpt}. Centrality parameters $N_{part} = N_{bin} + 1$ and $\nu(n_s)$ are then also determined. The solid points are implied by the Glauber analysis of Ref.~\cite{aliceppbprod}

Figure~\ref{fig4} (third) shows the differential cross section $d\sigma/dN_{part}$  (points) determined by a Glauber analysis of \ppb\ centrality~\cite{aliceppbprod}. That analysis is based in part on the assumption that $n_x \propto N_{part}$ for charge multiplicity $n_x$ within a V0A detector. The measured probability distribution $dP/dn_x$ on $n_x$ is assumed to be equivalent to the differential cross section in the form $(1/\sigma_0)d\sigma/dn_x$ to assign fractional cross sections to event classes on $n_x$ as in Fig.~1 of Ref.~\cite{aliceppbprod}. The fourth panel shows the consequent $N_{part}$ vs $\bar \rho_0$ relation (points). The TCM equivalent derived from \mmpt\ data is the solid curve. There is clearly a  fundamental disagreement between the two methods, prompting a reexamination of the Glauber model as it is applied to \pa\  collisions.

\section{$\bf p$-N exclusivity within p-Pb collisions}

Reference~\cite{exclusive} reports a possible resolution of the conflict between Glauber model and TCM noted above. The critical issue lies with the eikonal approximation, assumed within the Glauber model, as it applies to \pn\ collisions within a \ppb\ collision. The eikonal approximation applied to \pp\ collisions assumes that each participant parton in a projectile proton flies freely and may interact with any participant parton within a target proton that is inside its {\em eikonal corridor} defined by a parton-parton cross section. As a consequence of that assumption the number of binary parton-parton collisions (i.e.\ dijet production) in a simulated \pp\ collision goes as the 4/3 power of the number of participant partons, equivalent to \aa\ collisions with $N_{bin} \approx (N_{part}/2)^{4/3}$. The eikonal result is shown as the dashed curve in Fig.~\ref{fig1} (second) with $n_h/n_s \propto \bar \rho_s^{1/3}$. That dependence arises because in each event participants are restricted to a \pp\ overlap region determined by an impact parameter: i.e.\ \pp\ centrality matters.  In contrast,  \pp\ spectrum data follow a quadratic $N_{bin} \propto N_{part}^2$ relation with $n_h/n_s \propto \bar \rho_s$ characteristic of full overlap between \pp\ collision partners in each event. To resolve the conflict of centrality modeling for \ppb\ collisions an {\em exclusion-time constraint} may be imposed: a projectile proton already fully engaged in a \pn\ collision is excluded from interacting with the next nucleon for a time comparable to a nucleon diameter --  $\delta t \approx 1.5$ fm/c.

\begin{figure}[h]
	\includegraphics[width=1.24in]{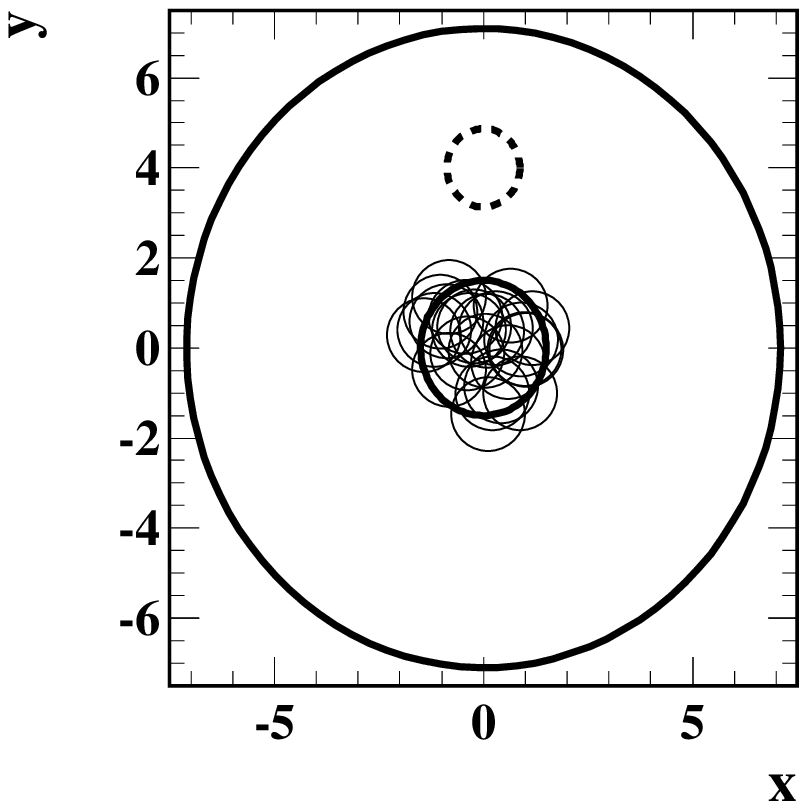}
	\includegraphics[width=1.24in]{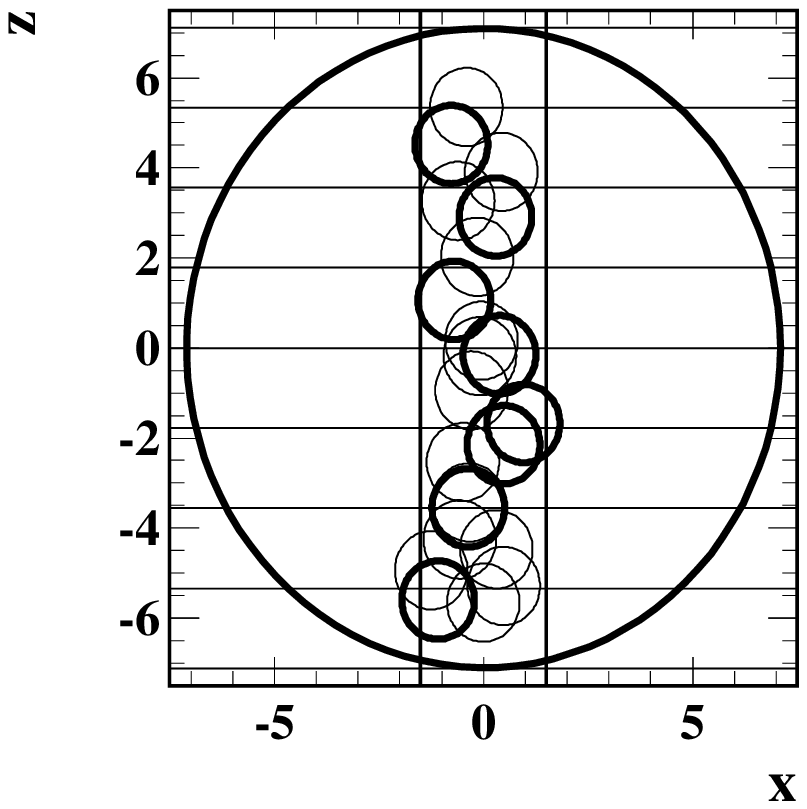}
	\includegraphics[width=1.26in]{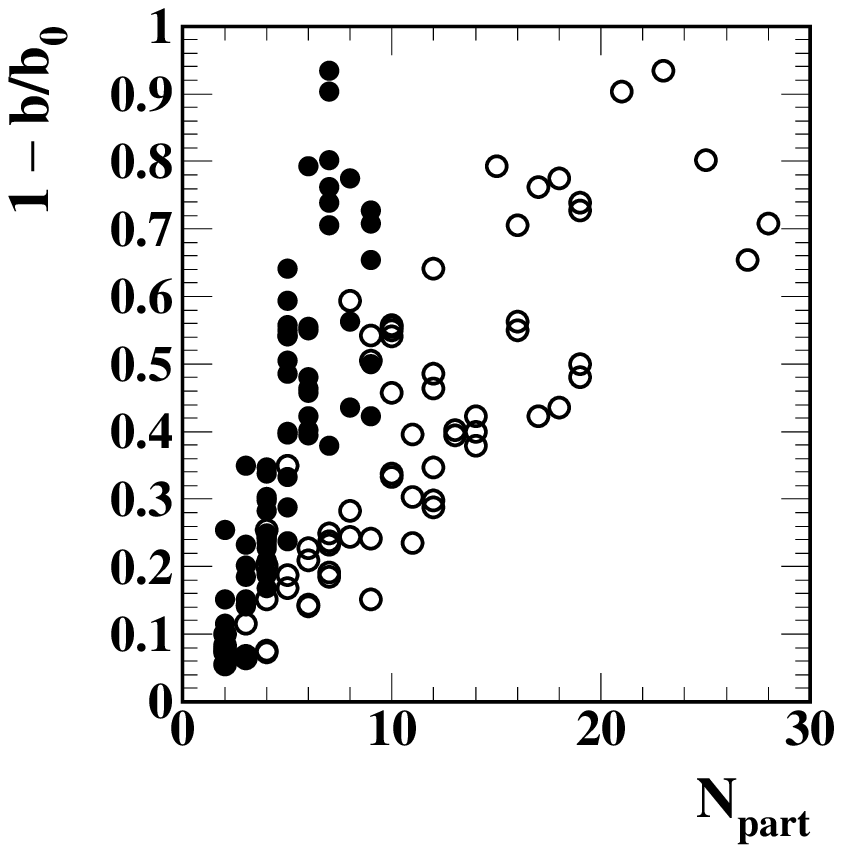}
	\includegraphics[width=1.26in]{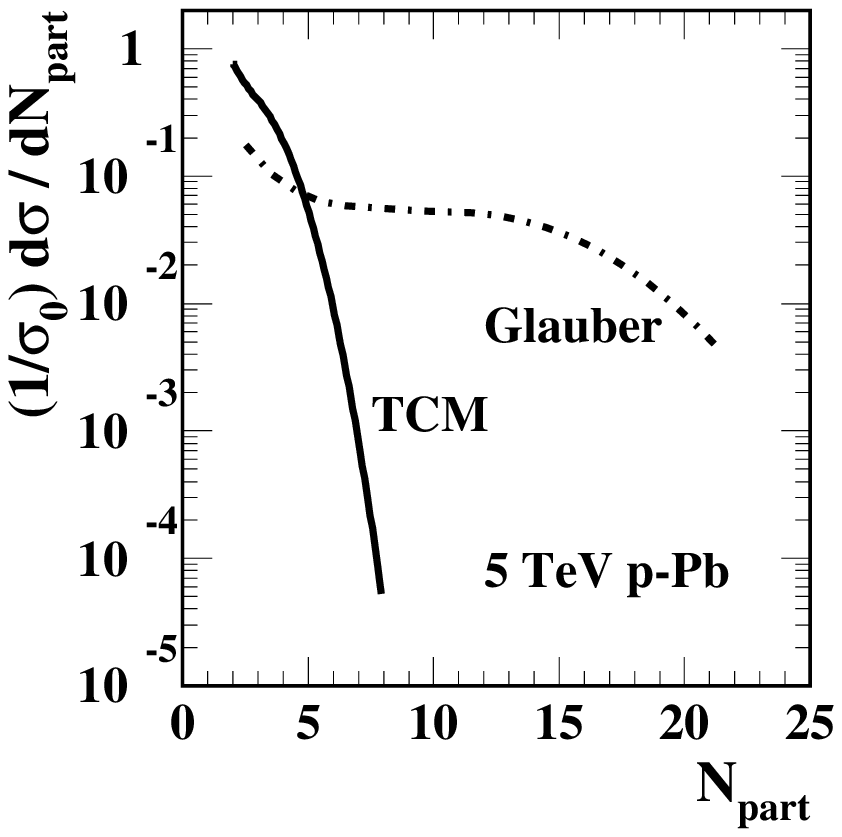}
	\caption{\label{fig5}
		First: Nucleons within eikonal corridor for $b = 0$ projectile.
		Second: \pn\ encounters (22 total circles) and actual collisions with exclusion-time constraint (8 dark circles).
		Third: Glauber-model relation of relative impact parameter $b/b_0$ to participant number $N_{part}$ with (solid) and without (open) an exclusion-time constraint.
		Fourth: Differential cross section on $N_{part}$ with (solid) and without (dash-dotted) an exclusion-time constraint.
	}   
\end{figure}

Figure~\ref{fig5} (first) shows a Glauber-model simulation of a \ppb\ collision viewed along the trajectory of a projectile proton with $b = 0$. The bold circle is the eikonal corridor defined by a nucleon-nucleon cross section. Any target nucleon with center inside the bold circle marks a \pn\ {\em geometric encounter}, of which there are 22 in this event (light circles). The second panel shows a transverse view of the collision system. The 22 encounters are again indicated by light circles. Imposing an exclusion-time constraint ($\delta t > 1.5$ fm/c between collisions) limits the actual \pn\ {\em collisions} to 8 (bold circles).

Figure~\ref{fig5} (third) shows number of participants vs relative impact parameter for an ensemble of simulated \ppb\ collisions with (solid points) and without (open circles) an exclusion-time constraint. The first and second panels represent one element of such an ensemble. The fourth panel shows the resulting cross-section distribution with exclusion-time constraint (solid curve) whereas the dash-dotted curve represents the Glauber data in Fig.~\ref{fig4} (third) with no constraint. Without the constraint a projectile nucleon may collide simultaneously with several overlapping target nucleons, leading to large overestimates of $N_{part}$ as reported in Ref.~\cite{tomglauber}. The need for an exclusion-time constraint within  \pa\ simulations is also consistent with the quadratic $\bar \rho_h \propto \bar \rho_s^2$ relation for isolated \pp\ collisions. That relation implies centrality is not relevant for such collisions -- any \pn\ {\em collision} exhibits 100\% overlap. Thus, individual \pn\ collisions are ``all or nothing'' and multiple {\em simultaneous} collisions are forbidden.

\section{p-Pb TCM and PID spectra} \label{sec-1}

The \ppb\ TCM for collision geometry, spectra and \mmpt\ data may be tested by application to spectrum data for identified (PID) hadrons reported in Ref.~\cite{aliceppbpid}. The TCM centrality derived from \mmpt\ data as described above is retained unchanged, as is the description of spectra and yields for unidentified hadrons. To describe data for identified hadrons new parameters $z_{si}$ and $z_{hi}$ represent soft- and hard-component yields of hadron species $i$ as {\em fractions} of the yields for unidentified hadrons. The TCM for normalized spectra from identified hadrons then follows from Eq.~(\ref{spectcm}) (second line)

\bea
\frac{\bar \rho_{0i}(y_t)}{\bar \rho_{si}} &=& \hat S_{0i}(y_t) + (z_{hi}/z_{si}) x(n_s) \nu(n_s) \hat H_{0i}(y_t)
\\ \nonumber
\frac{1}{\bar \rho_{si}} &=& \frac{1 + (z_{hi}/z_{si}) x(n_s) \nu(n_s)}{1+ x(n_s) \nu(n_s)} \frac{1}{z_{0i}} \frac{1}{\bar \rho_s},
\eea
where the second line defines the normalization for PID spectra in terms of the equivalent $1/\bar \rho_s$ for unidentified hadrons already determined in previous analysis. New parameters $z_h/z_s$ and $z_0$ are determined for each hadron species by the spectrum structure below $y_t = 2$ ($p_t \approx 0.5$ GeV/c) and are therefore unaffected by the spectrum hard component.

\begin{figure}[h]
	\includegraphics[width=2.5in]{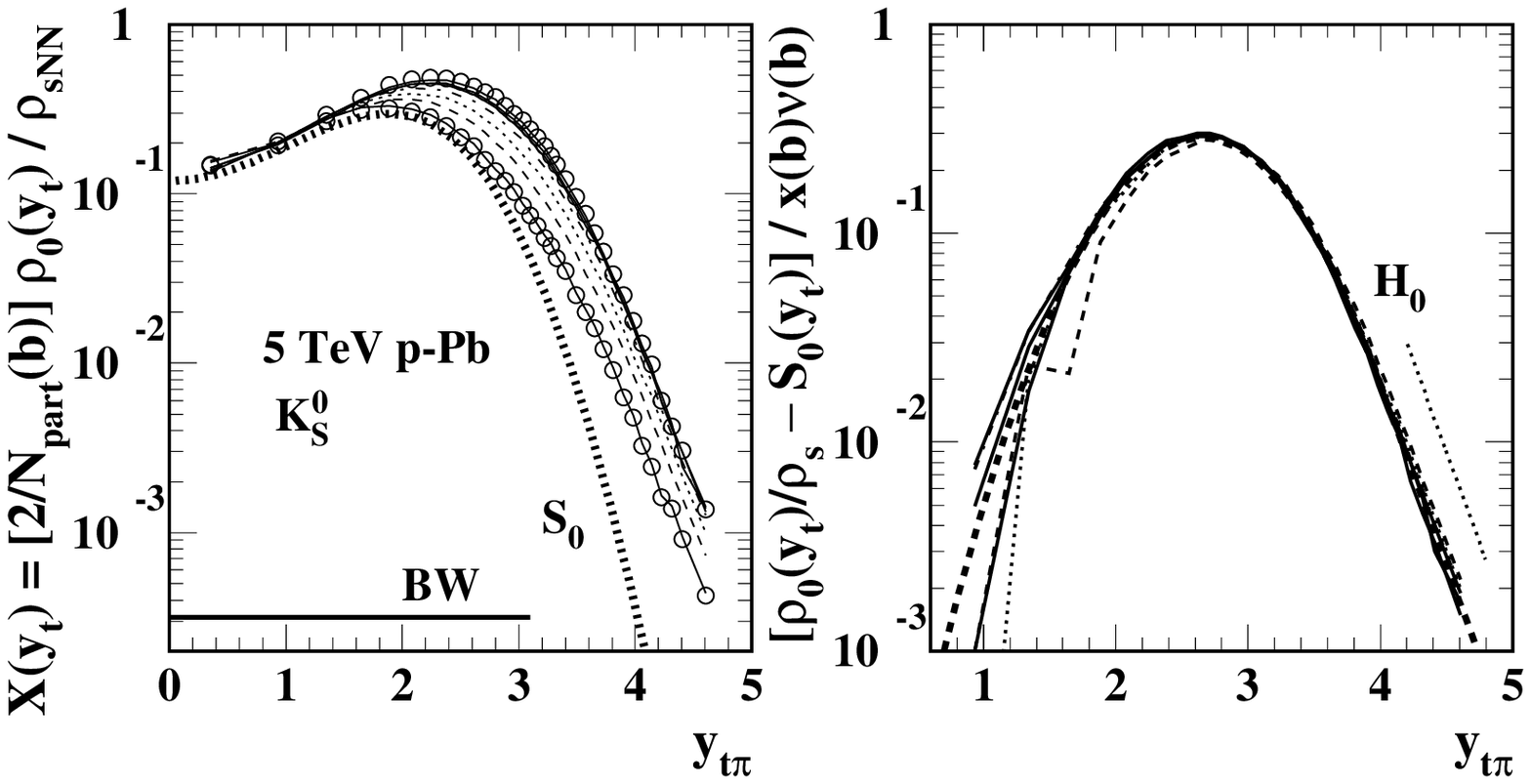}
	\includegraphics[width=2.5in]{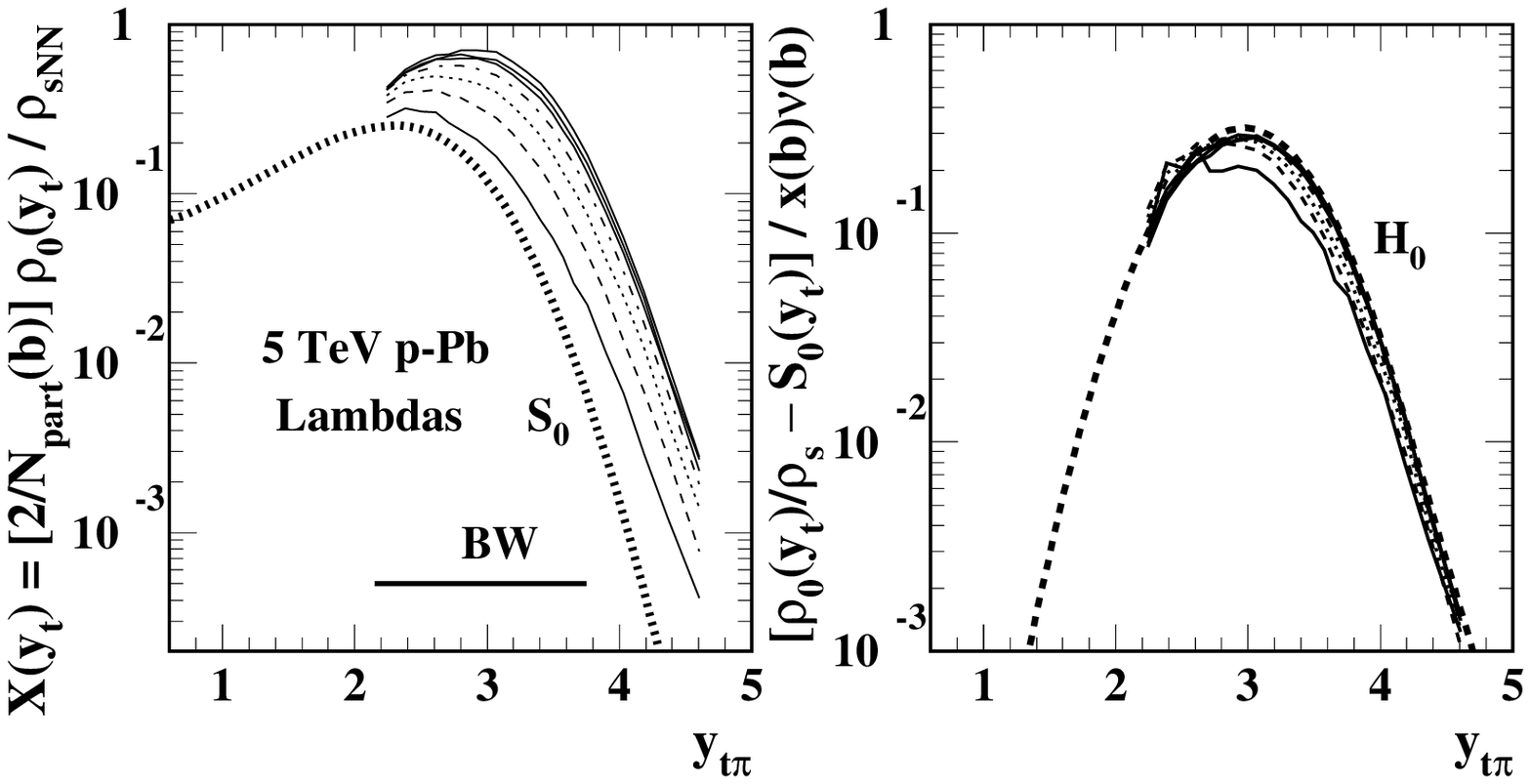}
	\caption{\label{fig6}
Spectra from seven centrality classes of 5 TeV \ppb\ collisions: Left: Full spectra (first) and isolated hard components (second) for $K^0_S$.
Right: Full spectra (first) and isolated hard components (second) for $\Lambda + \bar \Lambda$. The bold dotted and dashed curves are TCM models.
	}   
\end{figure}

Figure~\ref{fig6} shows full spectra (first and third) and extracted hard components (second and fourth) for $K_S^0$ and $\Lambda + \bar \Lambda$ from 5 TeV \ppb\ collisions. Similar results for pions, protons and charged kaons are obtained. The $K_S^0$ data below $y_t = 2$ ($p_t < 0.5$ GeV/c) confirm that the soft component, including slope parameter $T$, is independent of centrality. For each hadron species parameter $z_h/z_s$ is adjusted to bring spectra for seven centralities into coincidence for $y_t < 2$, and parameter $z_0$ is then adjusted to bring all spectra into coincidence with the unit-normal $\hat S_0(y_t)$ model function (bold dotted). Model parameters for charged and neutral kaons are assumed to be identical (consistent with spectrum data), and those for protons and Lambdas are similar.

\begin{figure}[h]
	\includegraphics[width=1.25in,height=1.25in]{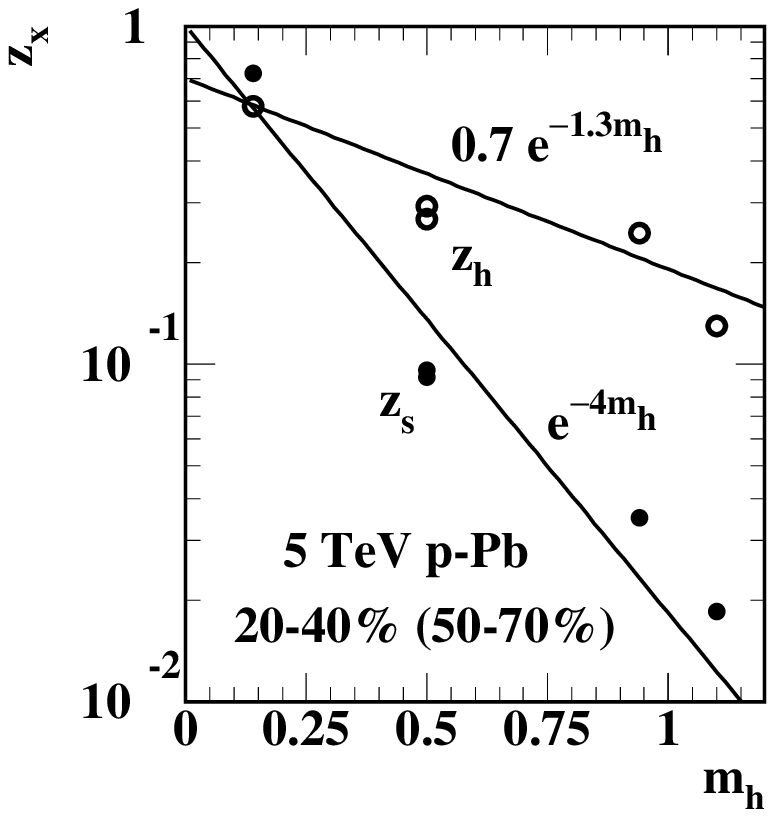}
	\includegraphics[width=1.25in]{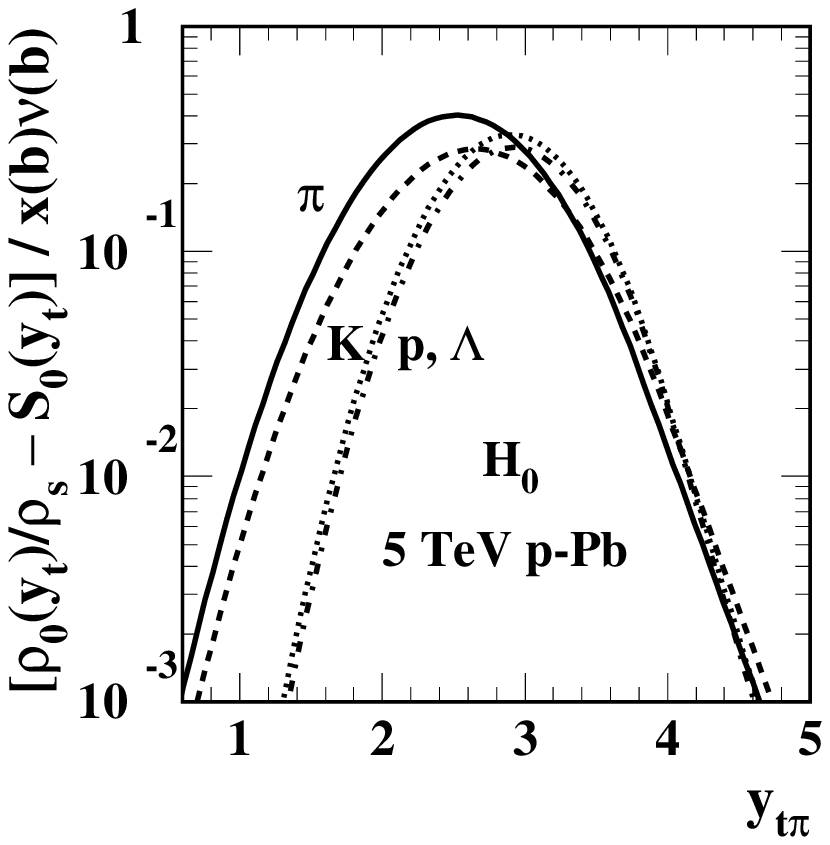}
	\includegraphics[width=1.25in]{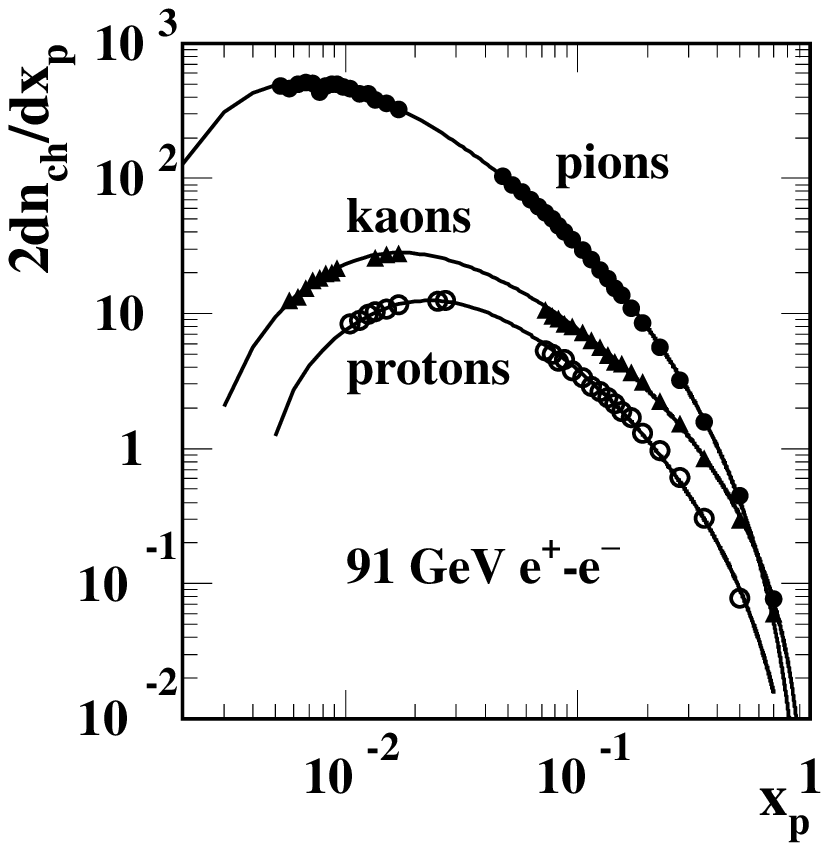}
	\includegraphics[width=1.25in,height=1.25in]{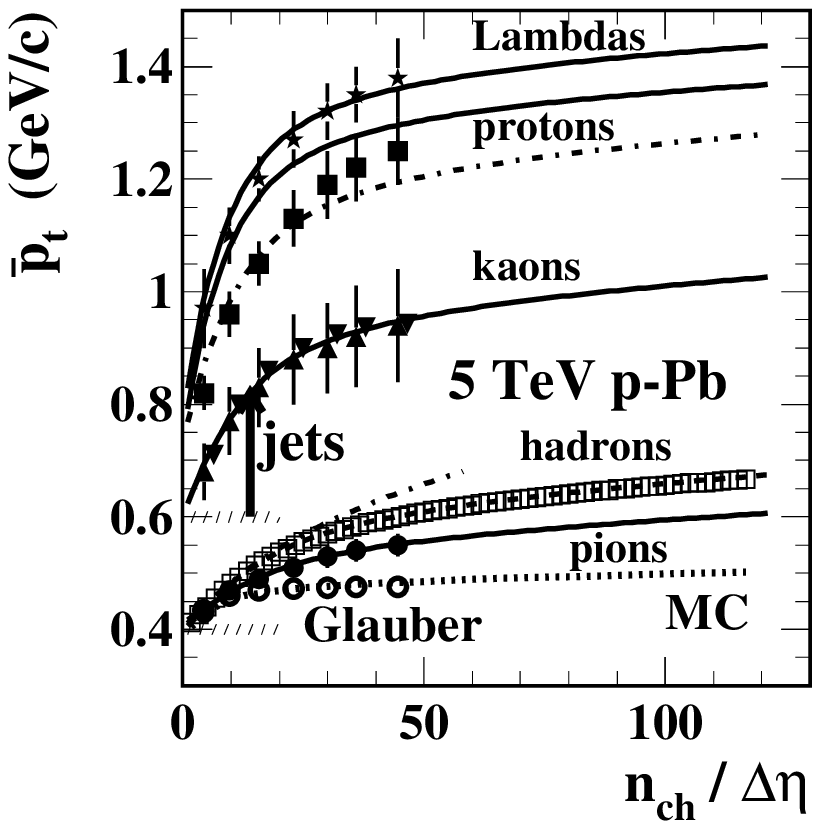}
	\caption{\label{fig7}
		First: TCM soft $z_s$ and hard $z_h$ fractions for identified hadrons relative to unidentified hadrons vs hadron mass.
		Second: Spectrum hard components for five hadron species.
		Third: Fragmentation functions for three hadron species.
		Fourth: PID \mmpt\ data (points) from 5 TeV \ppb\ collisions for seven multiplicity classes vs TCM  (solid curves).
	}   
\end{figure}

Figure~\ref{fig7} (first) shows $z_s$ and $z_h$ parameter values vs hadron mass for a specific \ppb\ centrality. The centrality numbers in parentheses are from Ref.~\cite{tomglauber} based on the \ppb\ TCM whereas the others are from Ref.~\cite{aliceppbprod} based on a Glauber analysis. The lines show exponential dependences on hadron mass. Whereas $z_s$ decreases strongly with mass as expected from the statistical model $z_h$ decreases much less quickly so that for baryons there is a large excess of jet fragments. The effect is most notable in Fig.~\ref{fig6} (third) where the Lambda hard component peaks near $y_t = 3$ ($p_t \approx 1.5$ GeV/c) and for more-central collisions dominates the soft component (bold dotted). Thus, the origin of the baryon/meson ``puzzle'' encountered in \aa\ collisions is due to jet production according to these data.

Figure~\ref{fig7} (second) shows TCM hard-component models $\hat H_0(y_t)$ for pions, kaons, protons and Lambdas [see bold dashed curves in Fig.~6 (second and fourth)]. The curves tend to coincide on the high-\yt\ side but have a strong mass dependence on the low-\yt\ side. It is interesting to compare those TCM hard components with corresponding fragmentation functions from LEP \ee\ collisions in the third panel (ALEPH data). The same mass trend is observed, buttressing the conclusion that the PID TCM hard components represent fragments from a common underlying jet spectrum.

Figure~\ref{fig7} (fourth) shows PID \mmpt\ vs \nch\ data for seven centrality classes of 5 TeV \ppb\ collisions from Ref.~\cite{aliceppbpid} (solid points) and corresponding TCM derived from the PID spectrum TCM described above (solid curves). The open circles are implied by results of the Glauber analysis of Ref.~\cite{aliceppbprod}, and the dotted curve (MC) represents default PYTHIA. \mmpt\ data for unidentified hadrons from Ref.~\cite{alicempt} (open squares) are included for comparison. Whereas the most central (0-5\%) point for PID data is reported to be at $\bar \rho_0 \approx 45$ the unidentified-hadron data from the same collaboration extend to 2.6 times that number, further evidence that the \ppb\ Glauber centralities reported in Ref.~\cite{aliceppbprod} can be questioned~\cite{tomglauber}. The TCM description of pion, kaon and Lambda \mmpt\ data is quite accurate, reflecting the quality of the TCM description of spectrum data. However, the proton data deviate substantially from the TCM expectation (proton solid curve). The difference in data is an apparent suppression of the proton hard component near its mode. The dash-dotted curve corresponds to inclusion of a fixed suppression factor to model the proton spectrum data. The remaining systematic deviations correspond to shifts of the proton hard component to higher \pt\ with increasing \nch\ or \ppb\ centrality. Similar but smaller shifts are  observed for Lambda data as in Fig.~\ref{fig6} (fourth panel) but not for kaons (second panel) or pions.

\section{Summary} \label{sec-1}

In this presentation the PYTHIA Monte Carlo (PMC) is confronted with several results from two-component model (TCM) analysis of particle data that challenge certain of its fundamental assumptions. The TCM accurately describes a broad array of data from \pp, \pa\ and \aa\ collision systems, and from SPS to top LHC collision energies, with only a few parameters. Physical interpretation of TCM soft and hard components is consistent with basic QCD energy trends and jet measurements.

The PMC is based in part on the assumption that almost all hadron production in \pp\ collisions arises from multiple parton interactions (MPIs) in each collision event that are described by pQCD. The underlying jet spectrum is extended down to $p_t = 0$ with a soft cutoff adjusted to accommodate data. In contrast, the TCM describes the same particle data via a soft component representing the majority of hadrons and a minority hard component representing jet production that is quantitatively compatible with measured jet spectra and fragmentation functions. The rate of MB dijet production per NSD \pp\ collision is $O(0.1)$, not multiple jets per collision. Measured \pt\ spectrum hard components are consistent with a jet spectrum cutoff near 3 GeV.

The {\em underlying event} (UE) is said to be accessed via charge multiplicity $N_\perp$ integrated within a limited azimuth interval (TR) relative to an imposed jet trigger. Increase of $N_\perp$ to a saturation value (pedestal) with increasing trigger \pt\ condition reportedly arises from increasing \pp\ centrality in response to requirement of more jets at higher \pt. But the pedestal effect is actually predicted by a simple running integral of the single-particle \pt\ spectrum, and the $N_\perp$ \pt\ spectrum indicates no change in the lower-\pt\ soft component (the actual UE) with trigger but substantial changes at higher \pt\ consistent with the triggered jet making a strong contribution to the TR.

\pp\ centrality is treated within the PMC by a Glauber model based on the eikonal approximation. As in \aa\ collisions dijet production then varies as the 4/3 power of the participants (i.e.
 low-$x$ gluons), as reflected in default-PMC predictions for \mmpt\ vs \nch. However, the TCM description of \pp\ spectra demonstrates a quadratic relation between MB jet production and participant gluons implying that centrality is irrelevant for \pp\ collisions, that each \pp\ collision involves full overlap of the collision partners and no restriction to an eikonal corridor for participant partons. The Glauber model applied to \ppb\ collisions also fails and for closely-related reasons: a \pn\ collision is ``all-or-nothing'' -- simultaneous collisions via partial overlaps are forbidden.

In response to failure of the default PMC to describe \mmpt\ vs \nch\ trends a color-reconnection or CR mechanism was introduced that minimizes string lengths (i.e.\ jet fragments) within MPI production. But that mechanism must be equivalent to strong changes in parton fragmentation with increasing jet number which are not observed in data. For \nch\ variation equivalent to 100-fold increase in dijet production spectrum hard components, equivalent to a convolution of jet spectrum and fragmentation-function ensemble, show little variation.

Successful TCM descriptions of \mmpt\ vs \nch\ data for \pp\ and \ppb\ collisions argue against recent claims for ``collectivity'' or flows in small collision systems. Strong increases in \mmpt\ are fully accounted for by the systematics of the spectrum hard component that is quantitatively related to MB dijets. Given a flow hypothesis it is difficult to account for the strong {\em decrease} in \mmpt\ with increasing system size from \pp\ to \pa\ to \aa, whereas in a TCM context dijet production per \nn\ pair depends quadratically on \nn\ \nch, which is largest in high-multiplicity \pp\ collisions and much smaller in central \ppb\ collisions. Given the accurate and complete description of \pt\ spectra in terms of a {\em fixed} soft component and a hard component quantitatively linked to jet production it is unlikely that radial flow (for example) plays a significant role in \pp\ or \pa\ collisions.

The ability of the TCM to accommodate a variety of A-B data is demonstrated by its extension to describe identified-hadron (PID) spectra from \ppb\ collisions. The TCM for unidentified hadrons from \ppb\ collisions, itself a very simple extension from \pp\ collisions, is maintained unchanged, and only two additional parameters (soft and hard fractions of unidentified-hadron yields) are required to describe PID data accurately. The same centrality estimates, based on \nn\ exclusivity, are also retained. The TCM description of PID data demonstrates that strong \mmpt\ increase with hadron mass, interpreted by some to be a manifestation of radial flow, is a consequence of MB dijet production: jet formation favors baryon production by a large factor, a result directly related to the so-called baryon/meson puzzle emerging from \aa\ collision data.

In conclusion, accurate TCM descriptions of a variety of A-B collision data emphasize the central role of MB dijets for hadron production near midrapidity, the absence of centrality dependence for \nn\ collisions and the importance of \pn\ ``exclusivity'' within \pa\ collisions. TCM results present strong challenges to several assumptions underlying the PYTHIA Monte Carlo, especially the prevalence of MPIs, relevance of the CR mechanism and conventional interpretations of underlying-event trends.


\end{document}